\documentstyle[aps,epsf,multicol,prl]{revtex}
\begin{document}
\draft

\author{J. Reichert$^1$, R. Ochs$^1$, D. Beckmann$^1$, H. B. Weber$^1$, M. Mayor$^1$, and H. v. L\"ohneysen$^{2,3}$}
\address{$^1$Forschungszentrum Karlsruhe, Institut f\"ur Nanotechnologie, D-76021 Karlsruhe}
\address{$^2$Forschungszentrum Karlsruhe, Institut f\"ur Festk\"orperphysik, D-76021 Karlsruhe}
\address{$^3$Physikalisches Institut, Universit\"at Karlsruhe, D-76128 Karlsruhe}
\date{\today}
\sloppy
\title{Driving current through single organic molecules}
\maketitle
\begin{abstract}
We investigate electronic transport through two types of conjugated molecules. Mechanically controlled break-junctions are used to couple thiol endgroups of 
single molecules to two gold electrodes. Current-voltage characteristics (IVs) of the metal-molecule-metal system are observed.
These IVs reproduce the spatial symmetry of the molecules with respect to the direction of current flow. We hereby unambigously detect an intrinsic property of the molecule, and are able to distinguish the influence of both the molecule and the contact to the metal electrodes on the transport properties of the compound system.
\end{abstract}
\begin{multicols}{2}
\narrowtext
Electronic transport through molecules has been first described theoretically in the 1970s \cite{aviram,joachim}. Since then, numerous experiments have been 
made where electrical current was driven through single-layer molecular films between two metallic electrodes \cite{metzger,chen,fisher}. Transport through 
single or at most few molecules on a gold surface has been observed with scanning tunneling microscopes, where the tip serves as a counterelectrode 
\cite{datta,joachim2}. In the tunneling regime, the current-voltage characteristics (IV) reflects the electronic density of states in the molecule and the 
conductance depends very sensitively on the tip distance. Only few experiments, however, have been realized which target current through a single molecule 
while the connection to both electrodes is symmetrically realized by a well defined chemical bond, which allows mechanical stability of the junction even at 
room temperature \cite{reed,kergueris}. However, to identify the IVs observed in these experiments as arising from a current through indeed a single sample 
molecule, comparison with some theoretical assumptions is required concerning the conductance amplitude, the transport mechanisms and the electrochemical 
potential of the sample.  
The experiment described in this letter demonstrates clearly and without the necessity of any assumptions that we observe electronic transport through a single 
molecule and not an ensemble of molecules. This is achieved by comparing the IVs of spatially symmetric and asymmetric, but otherwise similar molecules. 
Further, an analysis of the IV data gives new qualitative insight concerning the crucial role of the molecule-metal contact.

The two types of organic molecules were designed specifically for the present experiment (cf. Fig. 1). Both consist of a rigid rod-like central section with 
additional thiol functions on both ends to form stable covalent bonds to gold electrodes. 
Details of the synthesis will be published elsewhere.
As the molecules are very similar, comparable electronic properties are expected. However, their main difference is their spatial symmetry. While the antracene 
derivative (in the following referred to as "symmetric molecule") has a symmetry plane perpendicular to the molecule's sulphur-to-sulphur axis, in the nitro 
acetyl amine derivative ("asymmetric molecule"), the mirror symmetry is absent. For the symmetric molecule the IVs may be expected to be symmetric with respect 
to voltage inversion, for the asymmetric molecule a current flowing in positive direction or in negative direction will not necessarily result in the same 
magnitude of the voltage drop along the molecule.

\begin{figure}[b]
\epsfxsize5cm
\centering\leavevmode\epsfbox{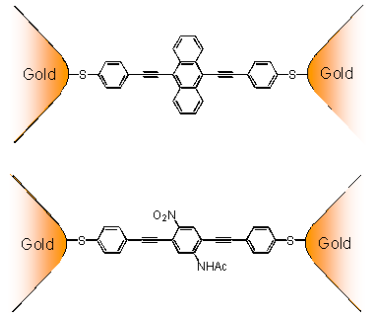}
 \caption{Scheme of the experimental setup: a spatially symmetric (9,10-Bis\-((2'-para-\-mercapto\-phenyl)-\-ethinyl)-\-anthra\-cene) and an asymmetric 
molecule (1,4-Bis\-((2'-\-para-\-mercapto\-phenyl)\-ethinyl)-2-acetyl-\-amino-5-\-nitro-\-benzene) in between two gold electrodes.}
\label{weber.fig1}
\end{figure}

 The length of both molecules is ~2 nm. To obtain a contact to a single molecule from both electrodes, an electrode pair with a distance matching exactly this 
length is required. We have chosen a lithographically fabricated mechanically-controlled break junction (MCB) to provide an electrode pair with tunable 
distance. The same technique was used in a previous experiment \cite{kergueris}. For more details on this technique see ref. \cite{vR}. A scanning electron 
microscope picture of a freshly prepared junction consisting essentially of a free standing Au bridge is shown in Fig. 2. This setup is mounted in a 
three-point support, designed to bend the substrate mechanically by means of a threaded rod which is driven by a DC motor. To prepare the experiment, we bend 
the substrate in order to elongate the bridge and finally it breaks. Then the two open ends form an electrode gap which can be adjusted mechanically with 
sub-\AA ngstrom precision.

\begin{figure}[tb]
\epsfxsize6cm
\centering\epsfbox{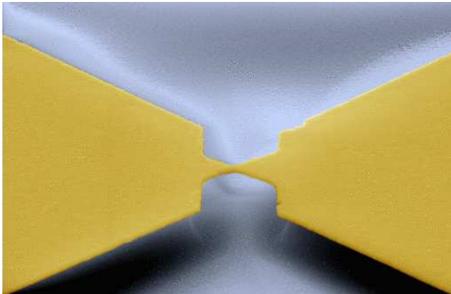}
\caption{Scanning electron microscope picture of the lithographically fabricated break junction. The setup consists of a metallic plate, covered by an 
insulating layer of polyimide. On top of this, a gold film with a small constriction (smallest diameter 50*50 nm$^2$) is deposited, laterally structured by 
e-beam lithography. Two electrodes lead outside to connect the bridge electrically. The polyimide is partially etched away so that in the constriction region, 
the bridge is freely suspended over the polyimide substrate.}
\label{weber.fig2}
\end{figure}

The molecules with acetyl protection groups at the ends are dissolved in tetrahydrofurane. A droplet of this solution is put on top of the opened MCB 
(electrode distance ~10 nm). The total exposure time is ~ 10-30 sec. When the molecules approach the surface of any of the gold electrodes, one of the acetyl 
protection groups splits off and a stable chemical bond between  the sulphur atom and the gold surface is established \cite{gryko}. The opposite side of the 
molecule remains protected at this stage. The coverage of the molecules on the gold surface is expected to be far below a completed monolayer, which would be 
formed only after hours. This is in contrast to previous experiments \cite{reed,kergueris}. Then the solvent is evaporated and the whole setup is mounted in an 
electromagnetically shielded box, which is pumped to a pressure of ~10$^{-7}$-10$^{-6}$ mbar. When the electrodes are approaching each other from large 
distances, the resistance decreases exponentially with distance, as expected for tunneling. At a certain distance, however, the system suddenly locks into a 
stable behaviour, which allows to record several IVs in the voltage range of [-1 V,1 V]. This stable configuration is interpreted as a metal-molecule-metal 
junction: when the first molecule touches the opposite Au surface, the second acetyl end-group is removed and a stable chemical bond is established from the 
single sample molecule to both electrodes. 
 
 \begin{figure}[tb]
\epsfxsize6cm
\centering\epsfbox{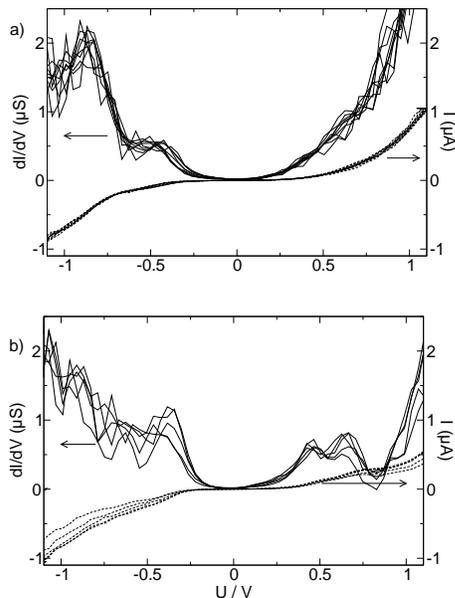}
 \caption{Transport data of the asymmetric molecule. a) Current-voltage (IV) raw data (dashed lines, nine subsequent voltage sweeps) on a stable junction and 
the numerically differentiated data $dI/dV$ (solid lines) from the above IV. b) Data from a subsequent junction.}
 \label{weber.fig3}
 \end{figure}

Fig. 3a shows nine IVs (dashed lines) obtained for a stable configuration with the asymmetric molecule. They are clearly nonlinear, displaying some rounded 
step-like features which appear presumably when transport through an additional molecular orbital is enabled by the bias voltage \cite{nitzan}. In addition, 
Coulomb blockade should be present in the system \cite{mujica,kemp}. Both effects were experimentally identified for example in tunneling through single 
semiconductor clusters \cite{banin}. Our data are highly reproducible as long as the junction remains stable. The current amplitude is about $0.7\mu$A at 1V. 
All observed stable junctions show currents in range of 0.2 -1 $\mu$A at 1 V. Beyond $V \approx 1.2$V, the current rises strongly and if higher voltages are 
applied, the junction becomes unstable.  Fig. 3a also displays the differential conductance $dI/dV$ (solid lines), obtained by numerical differentiation of the 
above current-voltage data. Here, the step-like features in the IV appear as peaks. The data are clearly asymmetric with respect to voltage inversion. Such IVs 
are stable within a time intervall ranging from 1 to 100 minutes. Thereafter, the system enters suddenly into a state of either considerably higher or lower 
conductance. When adjusting skillfully the electrode gap to a slightly different position, often another stable configuration can be established. Fig. 3b 
displays a data set obtained with the same MCB (and obviously an identical molecule) after the junction in Fig. 3a had become unstable. Compared to Fig. 3a, it 
shows similarities, but also differences. The first similarity is the total amplitude of the current, for example at $V=-1$V, of $0.7\mu$A. This current is 
barely sensitive to variations of the electrode distance, which indicates that indeed the current through a molecule is observed, with only a minor current 
contribution, if any, from direct metal-to-metal tunneling.

On the other hand, considerable differences can be seen: For example, two peaks at positive voltage appear in Fig. 3a, which are not visible in Fig. 3b. There 
are at least three reasons why differences may appear: First, different amounts of strain may have been applied to the molecule while establishing the contact. 
However, we do not believe that strain plays a dominant role in our experiments, because for stable contacts we can change the electrode distance by a few \AA 
ngstroms without altering the shape of the IV. Second, the proximity of neighbouring molecules/adsorbates or inhomogeneities of the electric fields may affect 
transport through the molecule. Inhomogeneous fields may influence spectroscopic properties of molecules only on energy scales in the mV range \cite{kohler}, 
yet only little is known about its influence on transport properties. Third, the atomistic structure of the contact region between molecule and electrode, 
which is different from junction to junction, may strongly influence the IVs. This is probably the dominant mechanism in our samples. An important indication 
is given by analyzing the first peak at negative voltage, which appears at $\approx450$mV in Fig. 3a and $\approx360$mV in Fig. 3b. This peak shift can be 
explained by different contact realisations in the following way: The gold-sulphur bond is a covalent, but strongly polarized bond. When the bond is 
established, a fraction of an electron charge is transferred onto the molecule due to the electronegativity of the S atom. Depending on the microscopic 
realisation of the bond (e.g. if the sulphur atom bonds to a gold atom at corners or kinks of the irregular surface etc), this charge transfer may vary and the 
molecule will be charged differently, which leads to an energy shift of the molecular orbitals. Hence, the peak belonging to the same molecular orbital will 
appear at different bias voltages, depending on the microscopic realisation of the contact. At first sight, one would expect by the same token that the 
conductance is very sensitive to contact variations too, as this is where the potential drops. This was in fact predicted theoretically for molecules of 
similar type \cite{yaliraki1}. Yet, the observed overall conductance seems not to be highly sensitive to these variations. We propose that our experimental 
procedure selects stable bonds on one side, which are limiting the conductance, while the opposite bond varies and governs the broadening as well as the charge 
transfer onto the molecule. The astonishing scenario that more stable bonds may yield lower conductance is in agreement with theoretical predictions, where the 
same tendency was observed for similar molecules \cite{yaliraki2}.

\begin{figure}[tb]
\epsfxsize6cm
\centering\epsfbox{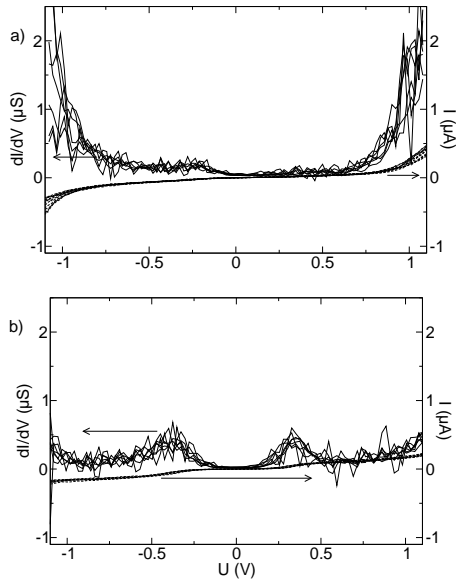}
 \caption{Current $I$ (dashed lines) and $dI/dV$ (solid lines) as a function of the bias voltage V from two subsequent junctions with the symmetric molecule.}
 \label{weber.fig4}
\end{figure}

Do we really observe individual molecules? Experimentally, the "lock-in" behaviour described above with a stable conductance not sensitive to the electrode 
distance (within $\pm 1$\AA) is a strong indication for a small discrete number of contributing molecules. The overall conductance does not vary by large 
amounts, therefore we deal always with approximately the same number of molecules. Indeed, when approaching the electrodes in a stable configuration further, 
we sometimes observe a discrete stable configuration where the conductance has roughly doubled, suggesting that we started out with a single molecule. Upon 
variation from one stable configuration to another peak positions of the differential conductance may shift as described above and their width may change, 
peaks may even appear or disappear altogether. In particular, we observe in a number of cases that the distinct asymmetry appears to be "mirrored" with respect 
to the bias voltage. This latter observation indicates clearly that we are not dealing with a large ensemble of molecules in parallel, where sample-to-sample 
fluctuations are averaged out, but rather with individual realisations of metal-molecule-metal junctions, which differ most probably in the microscopic 
arrangement of the contact. Our results, by exploring a number of different contact configurations, reflect the important effect of the environment on the 
transport properties of a single molecule. Note that sample-to-sample fluctuations are very familiar in single molecule spectroscopy in random media 
\cite{tamarat}. To conclude, the data reported thus far yield no definite proof, but strong indications that we observe electron transport through single 
molecules.

This is further confirmed by the data on symmetric molecules shown in Fig. 4a,b. The experimental procedure was the same as described above for the asymmetric 
molecule. The overall amplitude of the current $I \approx 0.6\mu$A at $V = 1$V  is similar to the values observed with the asymmetric molecule. At $V \approx 
±0.35$ V, a peak occurs in $dI/dV$ Fig. 4b. A similar peak in Fig. 4a is less pronounced and less symmetric. Apart from this slight asymmetry, the $dI/dV$ data 
look rather symmetric, in particular when compared with the strong asymmetry in $dI/dV$ of the asymmetric molecule (Fig. 3), with a very similar molecular 
structure along the sulphur-to-sulphur axis. More than 50\% of the stable IVs of the symmetric molecule were highly symmetric, while all of the IVs we observed 
with the asymmetric molecule were clearly asymmetric. This may be not very surprising, but allows a very important conclusion: what we measure is indeed the 
sample molecule and not an artefact caused by adsorbates or anything else. The IVs we observe reflect unambiguously an intrinsic property of the sample 
molecule: its spatial symmetry. Some of the IVs observed with the symmetric molecule, however, were asymmetric. An asymmetric IV with a symmetric molecule was 
previously observed in Ref. (9). For small asymmetries, there are several possible reasons: the symmetry can be broken by additional molecules, different 
electrode surfaces etc. To elucidate the question of more pronounced asymmetries as well, we manipulated intentionally a junction, starting with a highly 
symmetric IV. By carefully increasing the gap between the electrodes, we suddenly induced an asymmetric IV (Fig. 5a) with a steep current increase at positive 
bias and a rather small current at negative bias. Obviously, one of the two molecule-metal contacts was altered by the applied strain inducing an asymmetric 
IV. This IV was within minor deviations reproducible during three bias sweeps. We then played again with the electrode distance in both directions. The 
intention was now to move the strained bond to the other side. Two intermediate symmetric IVs were observed (Fig. 5b), similar to the initial configuration, 
which were then followed by asymmetric IVs, shown in Fig. 5c: Indeed, the steep current increase now occurs at negative bias and the current is smaller at 
positive bias, indicating a spatial inversion of the above contact configuration. This IV could be measured four times. Note that the molecule was not "lost"  
during this protocol, as no disruption of the IV measurement occurred. Hence, we apparently manipulated the contacts of only one molecule. This intentional 
manipulation of the IV symmetry gives further support that we observe the current through an individual molecule. In particular, the fact that asymmetric 
contact realisations can cause strong asymmetries in the IVs demonstrates the crucial importance of microscopic details within the contact region: although all 
contacts were chemically stable, different types of IVs could be observed with the same molecule. 

 \begin{figure}[tb]
\epsfxsize6cm
\centering\epsfbox{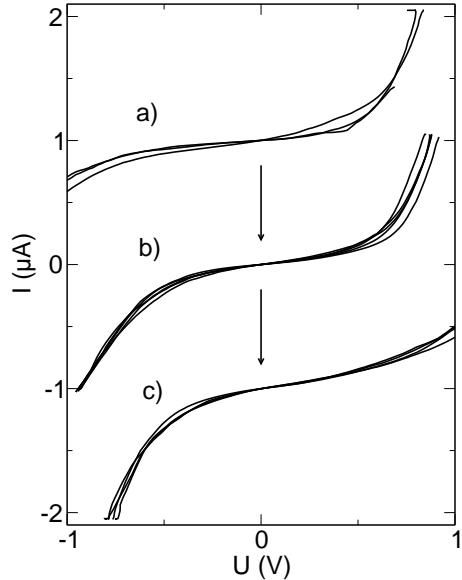}
 \caption{Three subsequent stable IVs with the symmetric molecule, observed during skillful manipulation of the electrodes. a) An asymmetry was mechanically induced. After an intermediate symmetric regime b), the IV appears inverted with respect to bias voltage (c). This proves the crucial influence of the 
microscopic contact arrangement and suggests the identification as an indiviual molecule. Note, that the junction was not lost  during this procedure. Data 
a) and c) are offset for visibility by $+1\mu$A and $-1\mu$A, respectively.}
 \label{weber.fig5}
 \end{figure}

In conclusion, we have performed conductance measurements through a self-assembled metal-molecule-metal junction. By comparison of spatially symmetric and 
asymmetric but otherwise similar organic molecules, we unambiguously identify an intrinsic property of the molecule: its symmetry. The body of data strongly 
suggest that individual molecules are observed, in particular a lock-in in stable configurations can be observed. This is further corroborated by the 
observation of sample-to-sample fluctuations which demonstrate the crucial importance of the coupling to the "environment", i.e. the electrodes.


\end{multicols}
\end{document}